# A Study of Fractional Schrödinger Equation-composed via Jumarie fractional derivative

**Joydip Banerjee[1], Uttam Ghosh[2a], Susmita Sarkar[2b] and Shantanu Das[3]**
Uttar Buincha Kajal Hari Primary school, Fulia, Nadia, West Bengal, India
email- joydip1955banerjee@gmail.com
[2]Department of Applied Mathematics, University of Calcutta, Kolkata, India;
[2a]email : uttam_math@yahoo.co.in
[2b] email : susmita62@yahoo.co.in
[3] Reactor Control Division BARC Mumbai India
email : shantanu@barc.gov.in

## Abstract

One of the motivations for using fractional calculus in physical systems is due to fact that many times, in the space and time variables we are dealing which exhibit coarse-grained phenomena, meaning that infinitesimal quantities cannot be placed arbitrarily to zero-rather they are non-zero with a minimum length. Especially when we are dealing in microscopic to mesoscopic level of systems. Meaning if we denote $x$ the point in space and $t$ as point in time; then the differentials $dx$ (and $dt$) cannot be taken to limit zero, rather it has spread. A way to take this into account is to use infinitesimal quantities as $(\Delta x)^\alpha$ (and $(\Delta t)^\alpha$) with $0 < \alpha < 1$, which for very-very small $\Delta x$ (and $\Delta t$); that is trending towards zero, these 'fractional' differentials are greater that $\Delta x$ (and $\Delta t$). That is $(\Delta x)^\alpha > \Delta x$. This way defining the differentials-or rather fractional differentials makes us to use fractional derivatives in the study of dynamic systems. In fractional calculus the fractional order trigonometric functions play important role. The Mittag-Leffler function which plays important role in the field of fractional calculus; and the fractional order trigonometric functions are defined using this Mittag-Leffler function. In this paper we established the fractional order Schrödinger equation-composed via Jumarie fractional derivative; and its solution in terms of Mittag-Leffler function with complex arguments and derive some properties of the fractional Schrödinger equation that are studied for the case of particle in one dimensional infinite potential well.

## 1. Introduction

Fractional order derivative are being extensively used by the authors recently to study different natural process and physical phenomena [1-11, 18, 21, 25]. The mathematicians of this era are trying to re construct general form of calculus using modification of the classical order derivative-and generalize the same to arbitrary order. Riemann-Liouville [11] definition of fractional derivative admits non-zero value for fractional differentiation of a constant. This increases the complexity in calculation and also contradicts the basic properties of classical calculus. To overcome this problem Jumarie [9, 19, 21] modified Riemann-Liouville definition of fractional derivative to obtain zero for derivative of constant and this type of derivative is also applicable for continuous but non-differentiable functions. More of that Mittag-Leffler [10] function was introduced in classical sense but it has several applications in fractional calculus field. The Mittag-Leffler function with complex argument gives the fractional sine and cosine functions for fractional order investigation [9]. Ghosh *et al* [11-12, 20] discussed about the solutions of various types of linear fractional differential equation (composed via Jumarie fractional derivative) in terms of Mittag-Leffler functions. All the previous works leads to the general formalism for fractional calculus which meets classical calculus at certain boundary. On the other hand researchers are using different fractional differential equations only incorporating the fractional order in place of classical order but what will be the actual equation in fractional sense if we start using the basic fractional equations? This is a challenging task to mathematicians.

If we have a point in space call it $x$, it originates from differential $dx$, that is $\int_0^x dx = x$. Now with a differential $(dx)^\alpha$, with $0 < \alpha < 1$, we have $(dx)^\alpha > dx$, while we have $dx$ the infinitesimal differential approaches zero, gives a space transformed as $x^\alpha$; that is $\int_0^x (dx)^\alpha \sim x^\alpha$ [21, 23, 24]. So is the case with a point at time call it $t$. This is coarse graining in scale of observation, where we come across the fractal space time, where the normal classical differentials $dx$ and $dt$, cannot be taken arbitrarily to zero, and then the concept of classical differentiability is lost [22]. The fractional order $\alpha$ is related to roughness character of the space-time, the fractal dimension [23-24]. Here in this paper fractional differentiation of order $\alpha$ is used to study the dynamic systems defined by function $f(x^\alpha, t^\alpha)$. In [23-25] the demonstration of the fractional calculus on fractal subset of real line is Cantor –Set, and the order $\alpha$ is taken accordingly.

In this paper we have studied the formulation on fractional form of quantum mechanics using fractional calculus. We tried to find internal behaviour of quantum realm. For this purpose we have developed fractional Schrödinger equation and tried to understand the nature of quantum mechanics for fractal region. This formulation also leads to normal quantum mechanics at limiting condition. The nature of the solution changes depending upon various fractional orders of differentiation, which may lead underlying significance of quantum mechanics. We had to modify De-Broglie's and Plank's hypothesis in fractional sense such that they remain intact if limiting conditions are used.

This paper is divided into separate sections and sub-sections. In section-2.0 some definitions of fractional calculus are described. In section-3.0 we discuss about "original plane progressive wave". Section 4.0 is about derivation of fractional Schrödinger equation. The section-4.1 deals with solution of fractional Schrödinger equation. The section-4.2 is about time independent fractional Schrödinger equation and Hamiltonian. In section5.0 we developed 'equation of continuity'. In section5.1 we discuss about properties of fractional wave function. The section5.2 discusses on further study on fractional wave function, in section-5.3 it is about 'orthogonal' and 'normal' conditions of wave functions. The section-6.0 is about operators and expectation values. The section-7.0 to section7.1 we discuss simple application particles in one dimensional infinite potential well. In section7.2 we have graphical representation of fractional wave function. Next section 7.3 is the graphical representation of probability density and then in section 7.4 we discussed energy calculation. In appendix-we have defined fractional quantities used in the study.

## 2. Some Definition of fractional calculus

There are several definitions of fractional derivative. The most leading definitions are Riemann-Liouville fractional derivative [5] Jumarie modified fractional derivative [9].

### a) Riemann-Liouville fractional derivative

Riemann-Liouville (R-L) fractional derivative of a function $f(x)$ is defined as

$$_aD_x^\alpha f(x) = \frac{1}{\Gamma(-\alpha+m+1)} \left(\frac{d}{dx}\right)^{m+1} \int_a^x (x-\tau)^{m-\alpha} f(\tau) d\tau$$

Where $m \leq \alpha < m+1$, $m$ is positive-integer. Using this definition, for a non-zero constant function the fractional derivative is not-zero [6], this is contrary to classical calculus.

### b) Jumarie modified definition of fractional derivative

To get rid of the problem of R-L fractional derivative, Jumarie modified [9, 19, 21] the definition of fractional derivative, for a continuous (but not necessarily differentiable) function $f(x)$, in the range $0 \leq x \leq a$ such that is

$$f^{(\alpha)}(x) = {}_0^J D_x^\alpha f(x) = \begin{cases} \frac{1}{\Gamma(-\alpha)} \int_0^x (x-\xi)^{-\alpha-1} f(\xi) d\xi, & \alpha < 0 \\ \frac{1}{\Gamma(1-\alpha)} \frac{d}{dx} \int_0^x (x-\xi)^{-\alpha} \left(f(\xi) - f(0)\right) d\xi, & 0 < \alpha < 1, \\ \left(f^{(\alpha-n)}(x)\right)^{(n)}, & n \leq \alpha < n+1, \quad n \geq 1 \end{cases}$$

In Leibniz's classical sense the Jumarie fractional derivative is defined via fractional difference. Let $f : \mathbb{R} \to \mathbb{R}, x \to f(x)$, denote a continuous (but not necessary differentiable) function, and let $h > 0$ denote a constant infinitesimal step. Define a forward operator $E_h[f(x)] = f(x+h)$; then the fractional difference on the right and of order $\alpha$, $0 < \alpha < 1$ of $f(x)$ is

$$\Delta_+^{(\alpha)} f(x) = \left(E_h - 1\right)^\alpha f(x)$$

$$= \sum_{k=0}^\infty (-1)^k \left({}^\alpha C_k\right) f\left(x + (\alpha-k)h\right)$$

Where, ${}^\alpha C_k = \frac{\alpha!}{k!(\alpha-k)!}$ are the generalized binomial coefficients. Then the Jumarie fractional derivative is following

$$f_+^{(\alpha)}(x) = \lim_{h \downarrow 0} \frac{\Delta_+^{(\alpha)} [f(x) - f(0)]}{h^\alpha}$$

$$= \frac{d^\alpha f(x)}{dx^\alpha}$$

Similarly one can have left Jumarie derivative by defining backward shift operator. In this Jumarie definition we subtract the function value at the start point, from the function itself and then the fractional derivative is taken (in Riemann-Liouvelli sense). This offsetting makes the fractional derivative of constant function as zero, and gives several ease and conjugation with classical integer order calculus, especially regarding chain rule for fractional derivatives, fractional derivative of product of two functions etc [21].

### c) Some techniques of Jumarrie derivative

Consider a function $f[u(x)]$ which is not differentiable but fractionally differentiable. Jumarie suggested [21] three different ways depending upon the characteristics of function.

$$D_1^\alpha(f[u(x)]) = f_u^{(\alpha)}(u)(u_x')^\alpha$$

$$D_2^\alpha(f[u(x)]) = (f/u)^{1-\alpha}(f_u'(u))^\alpha u^\alpha(x)$$

$$D_3^\alpha(f[u(x)]) = (1-\alpha)! u^{\alpha-1} f_u^{(\alpha)}(u) u^\alpha(x)$$

### d) Mittag-Leffler function and Fractional trigonometric functions

Mittag-Leffler function [10] is defined as infinite series, as following

$$E_\alpha(z) = \sum_{k=0}^{\infty} \frac{z^k}{\Gamma(1+\alpha k)}, \quad (z \in \mathbb{C}, \operatorname{Re}(\alpha) > 0)$$

The above definition is the one parameter Mittag-Leffler function. For $\alpha = 1$, it is simple exponential function i.e. $E_1(z) = e^z$. The fractional sine and cosine functions [4] defined by Mittag-Leffler function are

$$\cos_\alpha(t^\alpha) = \frac{E_\alpha(it^\alpha) + E_\alpha(-it^\alpha)}{2} = \sum_{k=1}^{\infty}(-1)^k \frac{t^{2k\alpha}}{(2k\alpha)!} = \sum_{k=1}^{\infty}(-1)^k \frac{t^{2k\alpha}}{\Gamma(2k\alpha+1)}$$

$$\sin_\alpha(t^\alpha) = \frac{E_\alpha(it^\alpha) - E_\alpha(-it^\alpha)}{2i} = \sum_{k=1}^{\infty}(-1)^k \frac{t^{(2k+1)\alpha}}{(2k\alpha+\alpha)!} \sum_{k=1}^{\infty}(-1)^k \frac{t^{(2k+1)\alpha}}{\Gamma(2k\alpha+\alpha+1)}$$

One of the most important properties of Mittag-Leffler function [11] is $^J D_x^\alpha E_\alpha(ax^\alpha) = a E_\alpha(ax^\alpha)$, meaning that Jumarie fractional derivative of order $\alpha$ of Mittag-Leffler function of order $\alpha$ in scaled variable $x^\alpha$ is just returning the function itself. This is in conjugation to classical calculus similar to exponential function and very-useful in solving fractional differential equation composed with Jumarie fractional derivative.

## 3. Original plane progressive wave

Plane wave is a special kind of wave which does not change direction with the time evolution, and a progressive wave is not disturbed by any boundary condition [13]. Consider a plane progressive wave which is propagating in the positive *x* direction with a constant velocity *v*. The general form is $f(x,t) = f(x-vt)$ [13]. We considers here the fractional plane progressive wave in the following form

$$f(x,t) = f(x^\alpha - v_\alpha t^\alpha), \quad 0 < \alpha \leq 1 \tag{1}$$

Here $v_\alpha$ is fractional velocity. When *α* is tending to one, this plane progressive wave turn to one dimensional plane-wave. Thus the wave what we considered in equation (1) is a plane progressive wave in *α*-th order fractional wave, where the space and time axis are transformed to $x^\alpha$ and $t^\alpha$ respectively, and $0 < \alpha \leq 1$, the value of *α* is a fractional number. Thus the wave we considered is a fractional plane wave moving in *x*-direction. Now the Jumarrie type fractional derivative [9, 21] is used to find various physical quantity and physical properties of the corresponding wave. Let us define the operator $^J D_x^\alpha \equiv \frac{\partial^\alpha}{\partial x^\alpha}, ^J D_x^{2\alpha} \equiv \frac{\partial^{2\alpha}}{\partial x^{2\alpha}}$ and $^J D_t^\alpha \equiv \frac{\partial^\alpha}{\partial t^\alpha}, ^J D_t^{2\alpha} \equiv \frac{\partial^{2\alpha}}{\partial t^{2\alpha}}$.

Consider $f[u(x,t)] = f(x^\alpha - v_\alpha t^\alpha)$, $0 < \alpha \leq 1$. Where $u(x,t) = x^\alpha - v_\alpha t^\alpha$, $0 < \alpha \leq 1$

Now we choose the differential trick that is $D_3^\alpha(f[u(x)]) = (1-\alpha)! u^{\alpha-1} f_u^{(\alpha)}(u) u^\alpha(x)$. Here the number 3 defines the third trick and finally 3 is not used in the differential operators. Now $D_x^\alpha(f[u(x,t)]) = (1-\alpha)! u^{\alpha-1} f_u^{(\alpha)}(u) u^\alpha(x,t)$ We know from standard fractional derivative that

$$u_x^{(\alpha)} = D_x^\alpha u(x,t) = D_x^\alpha[x^\alpha - v_\alpha t^\alpha] = D_x^\alpha[x^\alpha] = \alpha! = \Gamma(\alpha+1)$$

Clearly

$$D_x^\alpha(f[u(x,t)]) = \alpha!(1-\alpha)! u^{\alpha-1} f_u^{(\alpha)}(u) \tag{2}$$

Similarly

$$D_t^\alpha(f[u(x,t)]) = v_\alpha \alpha!(1-\alpha)! u^{\alpha-1} f_u^{(\alpha)}(u) \tag{2a}$$

From equations (2) and (2a) we get

$$D_t^\alpha f[u(x)] = v_\alpha D_x^\alpha f[u(x)]$$

Operating $D_x^\alpha$ in both side,

$$D_x^\alpha D_t^\alpha f[u(x)] = v_\alpha D_x^\alpha D_x^\alpha f[u(x)] = v_\alpha D_x^{2\alpha} f[u(x)] \tag{2b}$$

Now operating $D_t^\alpha$ on both sides

$$D_t^\alpha D_t^\alpha f[u(x)] = D_t^{2\alpha} f[u(x)] = v_\alpha D_t^\alpha D_x^\alpha f[u(x)] \tag{2c}$$

Now using the theorem in Appendix-8 and combining equations (2b) and (2c) we get

$$D_x^{2\alpha} f[u(x)] = \frac{1}{v_\alpha^2} D_t^{2\alpha} f[u(x)]$$

$${}^J D_t^{2\alpha}\left[f(x^\alpha - v_\alpha t^\alpha)\right] = v_\alpha^2 \left({}^J D_x^{2\alpha}\left[f(x^\alpha - v_\alpha t^\alpha)\right]\right) \tag{3}$$

Equation (4) represents the fractional wave equation of $\alpha$ order. If $\alpha = 1$ the equation turns to one dimensional classical wave equation for the plane progressive wave.

**4) Solution of the wave equation**

Consider the solution of equation (3) is of the type $f(x,t) = g(x^\alpha) r(t^\alpha)$ using this in equation (3) we get

$$\frac{1}{g(x^\alpha)} D_x^{2\alpha} g(x^\alpha) = \frac{1}{v_\alpha^2} \frac{1}{r(t^\alpha)} D_t^{2\alpha} r(t^\alpha)$$

Implying

$$r(t^\alpha)D_x^{2\alpha}g(x^\alpha) = \frac{1}{v_\alpha^2}g(x^\alpha)D_t^{2\alpha}r(t^\alpha)$$

Left hand side is space dependent and right hand side is time dependent. Clearly we can equate this equation with a constant let $k_\alpha^2$

The space part of the equation is now

$$\frac{1}{g(x^\alpha)}D_x^{2\alpha}g(x^\alpha) = k_\alpha^2 \qquad \text{or} \qquad D_x^{2\alpha}g(x^\alpha) = k_\alpha^2 g(x^\alpha)$$

Solution of the equation [11] is $g(x^\alpha) = bE_\alpha(\pm ik_\alpha x^\alpha)$, $b$ is a constant. Similarly the time part of the equation is $D_t^{2\alpha}r(t^\alpha) = k_\alpha^2 v_\alpha^2 r(t^\alpha)$. The solution is $r(t^\alpha) = BE_\alpha(\pm i\omega_\alpha t^\alpha)$ where we put $\omega_\alpha = k_\alpha v_\alpha$ and $B$ is a constant. Thus the general solution is

$$f(x,t) = AE_\alpha(\pm ik_\alpha x^\alpha)E_\alpha(\pm i\omega_\alpha t^\alpha) \tag{4}$$

$A$ is a constant.

## 4. Derivation of fractional Schrödinger equation

Consider a particle of mass $m$ moving with velocity $v$. According to de Broglie hypothesis [14] there is a wave associated with every moving material particle. The mathematical form of de Broglie hypothesis is $p = \hbar k$. Here momentum of the particle is denoted by $p$ and $k$ is wave vector in one dimension; $\hbar$ is reduced plank constant. Now Plank's hypothesis [14] shows energy $\varepsilon$ of a particle in quantum level is proportional to angular frequency that is $\varepsilon = \hbar\omega$. With this context it is assumed that De-Broglie hypothesis and Plank's hypothesis are also valid in fractional $\alpha$–th order with a modified form

$$p_\alpha = \hbar_\alpha k_\alpha \tag{5}$$

$$\varepsilon_\alpha = \hbar_\alpha \omega_\alpha \tag{6}$$

It is clear that if $\alpha=1$ the equations (5) and (6) reduces to the original form of de Broglie and Plank hypothesis. Here $\hbar_\alpha$ is reduced plank constant of $\alpha$ order; $\hbar = \frac{h}{2\pi}$, and $h$ is plank constant. Here we defined $\omega_\alpha$ as fractional order angular frequency and $k_\alpha$ as fractional order wave vector.

The general solution for equation (4) is $u = Af(x^\alpha - v_\alpha t^\alpha), 0 < \alpha \leq 1$ where $A$ is a constant. To find the explicit form of the solution Mittag-Leffler [10] function is taken as a trial solution [11] as in classical differential equation we consider the exp($x$) as the trial solution [15]. Thus

$$f(x^\alpha, t^\alpha) = AE_\alpha(-ik_\alpha x)E_\alpha(i\omega_\alpha t^\alpha) \tag{7}$$

Here $E_\alpha(-ik_\alpha x^\alpha)$ and $E_\alpha(i\omega_\alpha t^\alpha)$ are Mittag-Leffler functions of one parameter in complex variable. This is a trial solution of the equation (4). Now fractional velocity $v_\alpha$ can be defined as $v_\alpha = \omega_\alpha / k_\alpha$. It is assumed here that the velocity $v_\alpha$ is constant. This is the velocity of particle as well as the group velocity of wave.

Considering the particle possess the constant momentum $p_\alpha$ and constant energy $\varepsilon_\alpha$, i.e. energy and momentum does not vary with the propagation of the wave in space and time. Using conditions of (5) and (6) in the solution of (7) it can be written

$$f(x^\alpha, t^\alpha) = A E_\alpha\left(-\frac{i}{\hbar_\alpha} p_\alpha x^\alpha\right) E_\alpha\left(\frac{i}{\hbar} \varepsilon_\alpha t^\alpha\right) \tag{8}$$

This particle has some physical properties hidden inside. To investigate them some operations must be needed. It must be verified that how this function changes with the variation of space and time. This variation may be measured by an operation say $\alpha$ order fractional differentiation. Differentiating partially with respect to $x$ of order $\alpha$ of the equation (8), we get the following

$$^J D_x^\alpha f(x^\alpha, t^\alpha) = A\left(\frac{ip_\alpha}{\hbar_\alpha}\right) E_\alpha\left(-\frac{i}{\hbar_\alpha} p_\alpha x^\alpha\right) E_\alpha\left(\frac{i}{\hbar} \varepsilon_\alpha t^\alpha\right)$$

$$^J D_x^\alpha f(x^\alpha, t^\alpha) = \frac{i}{\hbar_\alpha}\left(p_\alpha f(x^\alpha, t^\alpha)\right) \tag{9}$$

and doing it once again we get

$$^J D_x^{2\alpha} f(x^\alpha, t^\alpha) = -\frac{1}{\hbar_\alpha^2} p_\alpha^2 f(x^\alpha, t^\alpha) \tag{10}$$

We have used $^J_0 D_x^\alpha \left[E_\alpha(ax^\alpha)\right] = a E_\alpha(ax^\alpha)$ [11].

Let's define $p_\alpha^2 = 2^\alpha m_\alpha \varepsilon_{\alpha K}$, where $m_\alpha$ is mass (in fractional frame), and $\varepsilon_{\alpha K}$ is kinetic energy of fractional order $\alpha$. Then equation (10) can be written as $^J D_x^{2\alpha} f(x^\alpha, t^\alpha) = -\frac{1}{\hbar_\alpha^2}(2^\alpha m_\alpha \varepsilon_{\alpha K}) f(x^\alpha, t^\alpha)$. This implies the following

$$-\frac{\hbar_\alpha^2}{2^\alpha m_\alpha}\left(^J D_x^{2\alpha}\left[f(x^\alpha, t^\alpha)\right]\right) = \varepsilon_{\alpha K} f(x^\alpha, t^\alpha) \tag{11}$$

Now the variation of the function with time is studied. Thus repeating the above steps i.e. taking Jumarie fractional derivative of order $\alpha$ w.r.t. time, we have following

$$^J D_t^\alpha f(x^\alpha, t^\alpha) = -\frac{i}{\hbar_\alpha} \varepsilon_\alpha f(x^\alpha, t^\alpha) \tag{12}$$

Here $\varepsilon_\alpha$ is total energy of the system. From the conservation of energy in fractal space it can be written that, Total energy ($\varepsilon_\alpha$) = (kinetic energy $\varepsilon_{\alpha k}$) + (potential energy $V(x^\alpha, t^\alpha)$). Thus

$$\varepsilon_\alpha = \varepsilon_{\alpha K} + V(x^\alpha, t^\alpha) \tag{13}$$

Using this condition of equation (13) in equation (12) and combining equation (11) and (12) we get the following

$$-\frac{\hbar_\alpha^2}{2^\alpha m_\alpha}\left(^J D_x^{2\alpha}\left[f(x^\alpha,t^\alpha)\right]\right)=\left((\varepsilon_\alpha-V(x^\alpha,t^\alpha))\right)f(x^\alpha,t^\alpha)$$

by rearranging above we obtain the following (14)

$$-\frac{\hbar_\alpha^2}{2^\alpha m_\alpha}\left(^J D_x^{2\alpha} f(x^\alpha,t^\alpha)\right)+V(x^\alpha,t^\alpha)f(x^\alpha,t^\alpha)=i\hbar_\alpha\left(^J D_t^\alpha f(x^\alpha,t^\alpha)\right) \quad (14)$$

Where $f(x^\alpha,t^\alpha)=AE_\alpha\left(-\frac{i}{\hbar_\alpha}p_\alpha x^\alpha\right)E_\alpha\left(\frac{i}{\hbar}\varepsilon_\alpha t^\alpha\right)$

This is the fractional Schrödinger equation of $\alpha$-th order. At the limit $\alpha = 1$ the equation reduces to the Schrödinger equation in one dimension space and time. This equation has the solution which will lead to certain interesting physical properties.

## 4.1 Solution of fractional Schrödinger equation

The basic method of the solution of equation (14) is the method of separation of variables. Method of separation of variables is by assuming that the solution is identified as the product of two different functions $\Phi(x^\alpha)$ and $T(t^\alpha)$, where $\Phi(x^\alpha)$ depends on the space variable and $T(t^\alpha)$ depends on time variable. The function is $f(x^\alpha,t^\alpha)=\Phi(x^\alpha)T(t^\alpha)$. Here $\Phi(x^\alpha)$ is the spatial function i.e. solely dependent on transformed-space $x^\alpha$ and $T(t^\alpha)$ is another function which is only function of transformed-time $t^\alpha$. Substitute $f(x^\alpha,t^\alpha)=\Phi(x^\alpha)T(t^\alpha)$ in equation (14), we obtain the following

$$-\frac{\hbar_\alpha^2}{(2)^\alpha m_\alpha}\frac{1}{\Phi(x^\alpha)}\frac{d^{2\alpha}\left[\Phi(x^\alpha)\right]}{dx^{2\alpha}}+V(x^\alpha)\Phi(x^\alpha)=\frac{i\hbar_\alpha}{T(t^\alpha)}\frac{d^\alpha\left[T(t^\alpha)\right]}{dt^\alpha} \quad (15)$$

Left hand side of the equation is space dependent and right hand is time dependent. Thus, to satisfy the equation (15) both sides must be equal to some constant. Now on the left side of the equation has a fractional potential term. This has the dimension of fractional energy, that is $[ML^2T^{-2}]^\alpha=[M^\alpha L^{2\alpha}T^{-2\alpha}]$. Clearly the constant must have the dimension of fractional energy due to homogeneity of dimension. From the right hand side of the equation, the dimension analysis allows us to choose the unit of the constant $\varepsilon_\alpha$ as $(Joule)^\alpha$ for fractional values of $\alpha$. Also it is supported by equation (12) that this constant is $\varepsilon_\alpha$, this is fractional energy. Now equation (15) can be written as two different equations, one is solely time dependent and another is only dependent on space.

$$\frac{i\hbar_\alpha}{T(t^\alpha)}\frac{d^\alpha\left[T(t^\alpha)\right]}{dt^\alpha}=\varepsilon_\alpha \quad (16)$$

$$-\frac{\hbar_\alpha^2}{(2)^\alpha m_\alpha}\frac{1}{\Phi(x^\alpha)}\frac{d^{2\alpha}\left[\Phi(x^\alpha)\right]}{dx^{2\alpha}}+V(x^\alpha)=\varepsilon_\alpha \qquad (17)$$

Solution of equation of type (16) was found by Ghosh *et al* [11] using the Mittag-Leffler functions in the following form $T(t^\alpha) \approx E_\alpha\left(-\frac{i}{\hbar_\alpha}\varepsilon_\alpha t^\alpha\right)$. Thus the solution of the equation (15) is (omitting the integral constant) is

$$f(x^\alpha,t^\alpha)=\Psi_\alpha=\Phi(x^\alpha)E_\alpha\left(-\frac{i}{\hbar_\alpha}\varepsilon_\alpha t^\alpha\right) \qquad (18)$$

For $\alpha = 1$, i.e. in limiting case the solution (18) turns to the solution of one dimensional classical Schrödinger wave equation.

## 4.2 Time independent fractional Schrödinger equation and fractional Hamiltonian

The equation (17) has no time dependent solution as well as the equation has no effect with the variation of time. Thus the equation (17) can be rearranged as

$$-\frac{\hbar_\alpha^2}{(2)^\alpha m_\alpha}\frac{d^{2\alpha}\left[\Phi(x^\alpha)\right]}{dx^{2\alpha}}-\left(\varepsilon_\alpha-V(x^\alpha)\right)\Phi(x^\alpha)=0 \qquad (19)$$

This is the time independent Schrödinger equation. This equation is potential dependent. So it is not possible to solve the equation without knowing the character of potential function. But it can be confirmed that the solution has only space dependency. So this equation says about only the characteristic of the particle with the variation of space. This equation is energy equation. Thus Hamiltonian can be constructed with the analogy of Schrödinger's one dimensional quantum wave equation. The Hamiltonian in terms of non-integer order derivative is defined as

$$\hat{H}_\alpha=-\frac{\hbar_\alpha^2}{(2)^\alpha m_\alpha}\frac{d^{2\alpha}}{dx^{2\alpha}}-V(x^\alpha) \qquad (20)$$

Therefore the equation (19) can be written in terms of Hamiltonian as

$$\hat{H}_\alpha\Phi=\varepsilon_\alpha\Phi \qquad (21)$$

This equation is nothing but an Eigen equation with the Eigen value $\varepsilon_\alpha$. The Eigen function of the equation is $\Phi$. The Eigen function is the "information centre" of a particle. One can operate it in various ways to find the corresponding physical property. From equation (21) it is clear that the Hamiltonian is such an operator, doing the same job. The Hamiltonian gives the correct information about the energy of the particle.

## 5. Equation of continuity

Consider the Schrödinger equation previously derived in equation (14)

$$-\frac{\hbar_\alpha^2}{2^\alpha m_\alpha}\left(^J D_x^{2\alpha}\left[f(x^\alpha,t^\alpha)\right]\right)+V(x^\alpha,t^\alpha)f(x^\alpha,t^\alpha)=i\hbar_\alpha\left(^J D_t^\alpha\left[f(x^\alpha,t^\alpha)\right]\right)$$

Multiply the equation with the complex conjugate of the solution say $f^*(x^\alpha,t^\alpha)$ in right side of the equation and rewriting the equation as followed

$$-\frac{\hbar_\alpha^2}{2^\alpha m_\alpha}f^*(x^\alpha,t^\alpha)\left(^J D_x^{2\alpha}\left[f(x^\alpha,t^\alpha)\right]\right)+V(x^\alpha,t^\alpha)f^*(x^\alpha,t^\alpha)f(x^\alpha,t^\alpha)$$
$$=i\hbar_\alpha f^*(x^\alpha,t^\alpha)\left(^J D_t^\alpha\left[f(x^\alpha,t^\alpha)\right]\right) \quad (22)$$

Let's take complex conjugate of the equation (14) and multiply with the function $f$ in right side of the equation and the new equation is

$$-\frac{\hbar_\alpha^2}{2^\alpha m_\alpha}f(x^\alpha,t^\alpha)\left(^J D_x^{2\alpha}\left[f^*(x^\alpha,t^\alpha)\right]\right)+V(x^\alpha,t^\alpha)f(x^\alpha,t^\alpha)f^*(x^\alpha,t^\alpha)$$
$$=-i\hbar_\alpha f(x^\alpha,t^\alpha)\left(^J D_t^\alpha\left[f^*(x^\alpha,t^\alpha)\right]\right) \quad (23)$$

Subtracting equation (22) from equation (23) we have following (by dropping $x^\alpha, t^\alpha$)

$$-\frac{\hbar_\alpha^2}{2^\alpha m_\alpha}\left(f^*\left(^J D_x^{2\alpha}[f]\right)-f\left(^J D_x^{2\alpha}[f^*]\right)\right)=i\hbar_\alpha\left(f^*\left(^J D_t^\alpha[f]\right)+f\left(^J D_t^\alpha f^*\right)\right) \quad (24)$$

Now the equation can be rewrite in the following form

$$-\frac{\hbar_\alpha^2}{2^\alpha m_\alpha}\,^J D_x^\alpha\left[f^*\left(^J D_x^\alpha[f]\right)-f\left(^J D_x^\alpha[f^*]\right)\right]$$
$$=i\hbar_\alpha\left(f^*\left(^J D_t^\alpha[f]\right)+f\left(^J D_t^\alpha[f^*]\right)\right) \quad (25)$$

Let's define $\frac{i\hbar_\alpha}{2^\alpha m_\alpha}\left(f^*\,^J D_x^\alpha f - f\,^J D_x^\alpha f^*\right)=j_\alpha$ that is probability current density of α-th order and $f^*f=\rho_\alpha$ is probability density of α-th order. For $\alpha=1$ the fractal probability density $f^*f=\rho_\alpha$ turns to the one dimensional probability density.

Thus the equation (25) reduces to

$$^J D_x^\alpha[j_\alpha]=\,^J D_t^\alpha[\rho_\alpha] \quad (26)$$

This is the equation of continuity of α-th order in one dimension. If probability density $f^*f=\rho_\alpha$ is independent of time, right hand side is zero. Thus the left hand side is also equal to zero. This implies that the one dimensional variation of current density with space is zero. Physical significance of the fact is that there is no source or sink of probability current density. This is the condition of stationary state. To satisfy the above condition of $\rho_\alpha$ the solution must of type $f(x^\alpha, t^\alpha) = \Psi_\alpha = \Phi(x^\alpha) E_\alpha\left(-i\left(\varepsilon_\alpha t^\alpha / \hbar_\alpha\right)\right)$. This is the stationary state of $\alpha$-th order. For $\alpha = 1$ the state is same as of the one dimensional stationary state.

## 5.1 Properties of fractional wave function

For further investigation it is needed to characterize the basic properties of the solution of fractional Schrödinger equation.

a) The fractional wave function must be continuous and should be single valued. As the particle has physical existence, the fractional wave function of the particle must be continuous at every position of space and time. If the fractional wave function is not continuous for some position or time then the particle will vanish in the middle of its trajectory which is not possible at all. Fractional wave function must be single valued i.e. for every position of space time the property of the particle is unique.

b) The fractional wave function must be square integrable in fractional sense i.e.

$\int_a^b \Psi_\alpha \Psi_\alpha^* dx^\alpha < \infty$ in the region $a \leq x \leq b$

c) Linear Combination of solutions of the fractional Schrödinger wave equation itself is a solution of the system. Thus linear combination of the wave function is another wave function.

d) The fractional wave function must vanish at the boundary. If it is does not then, the boundary itself loses its significances. The boundaries seize the motion of particle to go further. As a result the particle has to stop at the boundary and consequently the fractional wave function vanishes. Here boundary means perfectly rigid boundary. If we have an analogy with a vibrating string bounded by two certain points, then we can get no amplitude on the two end points. Mathematically the condition may be described as

$\Psi(a) = \Psi(b) = 0$ if the wave is in the region $a \leq x \leq b$

e) The fractional Schrödinger equation suggests that the $\alpha$-order fractional derivative ${}^J D_x^\alpha \equiv \frac{\partial^\alpha}{\partial x^\alpha}$ of wave function is continuous and single valued.

f) The $\alpha$ order fractional derivative of the wave function must vanish at the boundary. If not, condition of stationary state will violate as suggested in the equation of continuity.

g) The wave function must be normalized; that signifies the existence of the particle is certainly measured within a boundary.

## 5.2 Further study on fractional wave function

The general solution of fractional wave equation is $f(x^\alpha, t^\alpha) = \Psi_\alpha = \Phi(x^\alpha) E_\alpha \left(-i\varepsilon_\alpha t^\alpha / \hbar_\alpha\right)$. Its complex conjugate is $\Psi^*_\alpha = \Phi^*(x^\alpha) E_\alpha \left(i\varepsilon_\alpha t^\alpha / \hbar_\alpha\right)$. Multiplying $\Psi_\alpha$ with $\Psi^*_\alpha$ we get $\Psi_\alpha \Psi^*_\alpha = \Phi(x^\alpha) \Phi^*(x^\alpha)$. This quantity is independent of time. We define this quantity as "existence intensity" and $\Psi$ or as existence amplitude. In a certain boundary the particle exists certainly. So it can be written in mathematical form. Let $\Psi$ is defined in the boundary $-\infty \leq x \leq +\infty$. Then $\int_{-\infty}^{+\infty} \Psi_\alpha \Psi^*_\alpha dx^\alpha = \text{constant}$. Note that the notation $\int f(x) dx^\alpha$ implies fractional integration that is $\int_{-\infty}^{x} f(x) dx^\alpha = \frac{1}{\Gamma(\alpha)} \int_{-\infty}^{x} (x-\xi)^{\alpha-1} f(\xi) d\xi$ with $\alpha > 0$. If the particle does not exist in the boundary the integration vanishes. Now we can define $\int_{-\infty}^{+\infty} \Psi_\alpha \Psi^*_\alpha dx^\alpha = 1$ if the particle exists certainly and $\int_{-\infty}^{+\infty} \Psi_\alpha \Psi^*_\alpha dx^\alpha = 0$ if the particle does not exist anywhere. Clearly existence parameter $\int_{-\infty}^{+\infty} \Psi_\alpha \Psi^*_\alpha dx^\alpha$ is such that the condition $0 \leq \int_{-\infty}^{+\infty} \Psi_\alpha \Psi^*_\alpha dx^\alpha \leq 1$, gets satisfied. Consider we have to find the information about existence over a certain region inside the boundary. Then the quantity $\int_{-a}^{+b} \Psi_\alpha \Psi^*_\alpha dx^\alpha = l$ should be less than 1. It defines that particle is not localised and it is convenient because the particle behaves like wave and a wave is not localized. If all this existence parameter or probability adds, the whole probability is unity. From the equations (9), (10), (12), (17) we found that wave function is Eigen function of various operators.

## 5.3 Orthogonal and normal conditions of wave functions

Two functions $F(x)$ and $G(x)$ defined in the region $a \leq x \leq b$ are orthogonal if their inner product is zero [15]. From the analogy of this orthogonal condition in $\{x\}$ space we can define the orthogonal condition for $\{x^\alpha\}$ space with following fractional integration operation such that,

$$\langle F | G \rangle = \int_a^b F^*(x^\alpha) G(x^\alpha) dx^\alpha = 0 \qquad (27)$$

Here $\langle F | G \rangle = \int_a^b F^*(x^\alpha) G(x^\alpha) dx^\alpha$ is defined as inner product of order $\alpha$ where $F^*(x^\alpha)$ is complex conjugate.

In the same way normal condition can be defined by the following fractional integration

$$\langle F | G \rangle = \int_a^b F^*(x^\alpha) G(x^\alpha) dx^\alpha = 1 \qquad (28)$$

The general solution of wave function is $\Psi_\alpha = \sum_i^\infty c_i \psi_i$, here $c_i$ is some constant. Here $i$ is dummy index. The complex conjugate of the solution is $\Psi_\alpha^* = \sum_j^\infty c_j^* \psi_j^*$ the inner product using Dirac's Bracket notation

$$\langle \Psi_\alpha | \Psi_\alpha \rangle = \sum_j^\infty \sum_i^\infty c_j^* c_i \langle \psi_{j\alpha}^* \psi_{i\alpha} \rangle \qquad (29)$$

From orthogonal and normal condition we have the following

$\langle \Psi_\alpha | \Psi_\alpha \rangle = \sum_j^\infty \sum_i^\infty c_j^* c_i \langle \psi_{j\alpha}^* \psi_{i\alpha} \rangle = 0$ if $i \neq j$ and $\langle \Psi_\alpha | \Psi_\alpha \rangle = \sum_j^\infty \sum_i^\infty c_j^* c_i \langle \psi_{j\alpha}^* \psi_{i\alpha} \rangle = 1$ if $i = j$. In the similar way $\langle \psi_{j\alpha} | \psi_{i\alpha} \rangle = 1$ if $i = j$. Clearly $\langle \Psi_\alpha | \Psi_\alpha \rangle = \sum_i^\infty c_i c_i^* = 1$. More precisely it can be written $\langle \Psi_\alpha | \Psi_\alpha \rangle = \sum_i^\infty |c_i|^2 = 1$. Now we can define $c_i$ as existence coefficient or probability coefficient.

## 6. Operators and expectation values

In quantum mechanics all the measureable quantities that cannot be measured directly are measured by expectation values [15]. So in the case of $\alpha$ order quantum mechanics it needs to define operators for every measurable quantity. For this purpose there must be some rules of choosing operators.

i) Every operator must be Eigen operator of the wave function.

ii) Eigen value of the operator defines measurable quantity.

iii) Expectation value of an operator is the measure of the corresponding operator.

Consider an operator $\hat{A}_\alpha$ operates on a certain function $\Psi_{i\alpha}$ such that

$$\hat{A}_\alpha \Psi_{i\alpha} = \lambda_i \Psi_{i\alpha} \qquad (30)$$

From the general form of $\Psi_\alpha$, the equation turns to $\hat{A}_\alpha [\psi_\alpha] = \sum_i^\infty c_i \hat{A}_\alpha \psi_{i\alpha} = \sum_i^\infty \lambda_i c_i \psi_{i\alpha}$. Thus $\lambda_i$ cannot be determined directly. For the correct information of the system we have to find the mean value or expectation value of the system. Expectation value of an operator is defined as

$$\langle A \rangle = \frac{\int_{-\infty}^{+\infty} \psi_{i\alpha} \hat{A}_\alpha \psi_{i\alpha}^* dx^\alpha}{\int_{-\infty}^{+\infty} \psi_{i\alpha} \psi_{i\alpha}^* dx^\alpha} \qquad (31)$$

For every physical measurable quantity there is corresponding expectation value.

## 7. Simple application-particles in one dimensional infinite potential well

Consider a particle is bounded by a one dimensional infinite potential well with length $x = 0$ to $x = a$ for $0 \leq x \leq a$. The potential is defined here is of the type of $V = 0$ if $0 \leq x \leq a$ and $V = \infty$ otherwise. Thus the particle is strictly bounded by the potential well in the transformed scale too. So wave function is also zero outside the well. For continuity, the wave function must vanish at the boundaries also i.e. $\Phi(0) = \Phi(a^\alpha) = 0$. The fractional Schrödinger equation as suggested in equation (19) is

$$-\frac{\hbar_\alpha^2}{(2)^\alpha m_\alpha} \frac{d^{2\alpha}\Phi(x^\alpha)}{dx^{2\alpha}} - \left(\varepsilon_\alpha - V(x^\alpha)\right)\Phi(x^\alpha) = 0$$

In this we take $V(x^\alpha) = 0$. Thus the equation is of the form as follows

$$-\frac{\hbar_\alpha^2}{(2)^\alpha m_\alpha} \frac{d^{2\alpha}\Phi(x^\alpha)}{dx^{2\alpha}} - \varepsilon_\alpha \Phi(x^\alpha) = 0$$

Rearranging we get the following

$$\frac{d^{2\alpha}\Phi(x^\alpha)}{dx^{2\alpha}} + \frac{(2)^\alpha m_\alpha \varepsilon_\alpha}{\hbar_\alpha^2} \Phi(x^\alpha) = 0$$

Let's take

$$\frac{(2)^\alpha m_\alpha \varepsilon_\alpha}{\hbar_\alpha^2} = k_\alpha^2 \qquad (31a)$$

and the equation is now is following

$$\frac{d^{2\alpha}\Phi(x^\alpha)}{dx^{2\alpha}} + k_\alpha^2 \Phi(x^\alpha) = 0 \qquad (32)$$

This equation has solution as suggested by Ghosh et al [11]

$$\Phi(x^\alpha) = A E_\alpha\left(-ik_\alpha x^\alpha\right) + B E_\alpha\left(ik_\alpha x^\alpha\right) \qquad (33)$$

Using boundary condition $\Phi(0) = \Phi(a^\alpha) = 0$, we get $A + B = 0$. Thus the solution (32) is $\Phi(x^\alpha) = B\left(E_\alpha\left(-ik_\alpha x^\alpha\right) - E_\alpha\left(ik_\alpha x^\alpha\right)\right)$

Using the definition of fractional sine function [9] we write the following

$$\Phi(x^\alpha) = C \sin_\alpha(k_\alpha x^\alpha) \qquad (34)$$

Using boundary condition on equation (34) we get again

$$\Phi(a^\alpha) = C \sin_\alpha(k_\alpha a^\alpha) = \Phi(0) = 0 \qquad (34a)$$

As defined by Jumarie [11] $\sin_\alpha(x^\alpha) = \sin_\alpha((x+M_\alpha)^\alpha)$ here we defined $M_\alpha$ as first order zero or first zero crossing [16] for $\alpha$-th ordered sin function. Since $\sin_\alpha(0) = 0$, therefore

$$\sin_\alpha((M_\alpha)^\alpha) = 0 \tag{34b}$$

Comparing equations (34a) and (34b) $\sin_\alpha(k_\alpha a^\alpha) = \sin_\alpha((M_\alpha)^\alpha)$, we get $ka^\alpha = (M_\alpha)^\alpha$ implying

$$k_\alpha = \left(\frac{M_\alpha}{a}\right)^\alpha \tag{34c}$$

Using the value of $k_\alpha = \left(\frac{M_\alpha}{a}\right)^\alpha$ in equation (34) the solution is $\Phi(x^\alpha) = C \sin_\alpha\left(\left(\frac{M_\alpha}{a}\right)^\alpha x^\alpha\right)$

### 7.1 Normalization of wave function

The normalization condition for wave function of $\alpha$-th order is

$$\int_0^a \Phi_\alpha \Phi_\alpha^* dx^\alpha = 1 \tag{35}$$

Now $\Phi(x^\alpha) = C \sin_\alpha(k_\alpha x^\alpha)$ is real so $\Phi_\alpha \Phi_\alpha^* = |\Phi_\alpha|^2$; then $\int_0^b |\Phi_\alpha|^2 dx^\alpha = 1$. Here $\Phi_\alpha \Phi_\alpha^* = |\Phi_\alpha|^2$ is $|\Phi(x^\alpha)|^2 = C^2 \sin_\alpha^2(k_\alpha x^\alpha)$. Thus the integration is $C^2 \int_0^a \sin_\alpha^2(k_\alpha x^\alpha) dx^\alpha = 1$. To integrate the equation an identity must be developed. By definition we have following

$$\cos_\alpha(2x^\alpha) = \frac{E_\alpha(2ix^\alpha) + E_\alpha(-2ix^\alpha)}{2}$$

Now we have following identities

$$\cos_\alpha(2x^\alpha) - 1 = \frac{E_\alpha(2ix^\alpha) + E_\alpha(-2ix^\alpha)}{2} - 1$$

$$\cos_\alpha(2x^\alpha) - 1 = \frac{E_\alpha(2ix^\alpha) + E_\alpha(-2ix^\alpha) - 2}{2}$$

$$\cos_\alpha(2x^\alpha) - 1 = \frac{\left(E_\alpha(ix^\alpha)\right)^2 + \left(E_\alpha(-ix^\alpha)\right)^2 - 2E_\alpha(ix^\alpha)E_\alpha(-ix^\alpha)}{2}$$

$$\cos_\alpha(2x^\alpha) - 1 = \frac{\left(E_\alpha(ix^\alpha) - E_\alpha(-ix^\alpha)\right)^2}{2}$$

By definition we have

$$\sin_\alpha(x^\alpha) = \frac{E_\alpha(ix^\alpha) - E_\alpha(-ix^\alpha)}{2i} \tag{36}$$

So we get the following identity

$$1 - \cos_\alpha(2x^\alpha) = 2\sin_\alpha^2(x^\alpha) \tag{37}$$

Using the identity of equation (37), we have following

$$C^2 \int_0^a \sin_\alpha^2(k_\alpha x^\alpha) dx^\alpha = 1$$

$$C^2 \left(\tfrac{1}{2}\right) \left( \int_0^a (1 - \cos_\alpha 2k_\alpha x^\alpha) dx^\alpha \right) = 1$$

$$C^2 \left(\tfrac{1}{2}\right) \int_0^a dx^\alpha - C^2 \left(\tfrac{1}{2}\right) \int_0^a \cos_\alpha(2k_\alpha x^\alpha) dx^\alpha = 1$$

$$\frac{C^2}{2} \left( \frac{x^\alpha}{\Gamma(1+\alpha)} - \frac{\sin_\alpha(2k_\alpha x^\alpha)}{2k_\alpha} \right)_{x=0}^{x=a} = 1$$

$$\frac{C^2}{2} \left( \frac{a^\alpha}{\Gamma(1+\alpha)} - \frac{\sin_\alpha(2k_\alpha a^\alpha)}{2} \right) = 1$$

Also $\frac{\sin_\alpha(2k_\alpha a^\alpha)}{2}$ is zero as suggested by boundary condition. Thus $\left(\frac{C^2}{2}\right)\left(\frac{a^\alpha}{\Gamma(1+\alpha)}\right) = 1$ or $C = \sqrt{\frac{2\Gamma(1+\alpha)}{a^\alpha}}$. Now the solution is $\Phi(x^\alpha) = \sqrt{\frac{2\Gamma(1+\alpha)}{a^\alpha}} \sin_\alpha\left(\left(\frac{M_\alpha}{a}\right)^\alpha x^\alpha\right)$. For $\alpha = 1$ the solution is converted to one dimensional solution for one dimensional Schrödinger equation of infinite potential well.

### 7.2 Graphical representation of wave function

Graphical presentation of $\Phi(x^\alpha) = \sqrt{\frac{2\Gamma(1+\alpha)}{a^\alpha}} \sin_\alpha(k_\alpha x^\alpha)$ for different values of the fractional order for $a = 10$ unit is shown in the figure-1. Before the plot we need to know the values of $M_\alpha$ for various $\alpha$. We found using Wolfram Mathematica-9 the various approximate values of $(M_\alpha)$ for 10000 terms of Mittag-Leffler sin function and numerically it can be shown that $\sin_\alpha(x^\alpha)$ losses periodicity for $\alpha < 1$. They are listed below in Table-1

**Table-1:** First zeros of function $\sin_\alpha(x^\alpha)$ after $x = 0$ for different $\alpha$.

| $\alpha$ | $(M_\alpha)$ |
|---|---|
| 0.736 | 3.19590 |
| 0.75 | 2.96354 |
| 0.80 | 2.80104 |
| 0.85 | 2.80556 |
| 0.90 | 2.87596 |
| 0.95 | 2.99051 |
| 1.0 | 3.14159 |

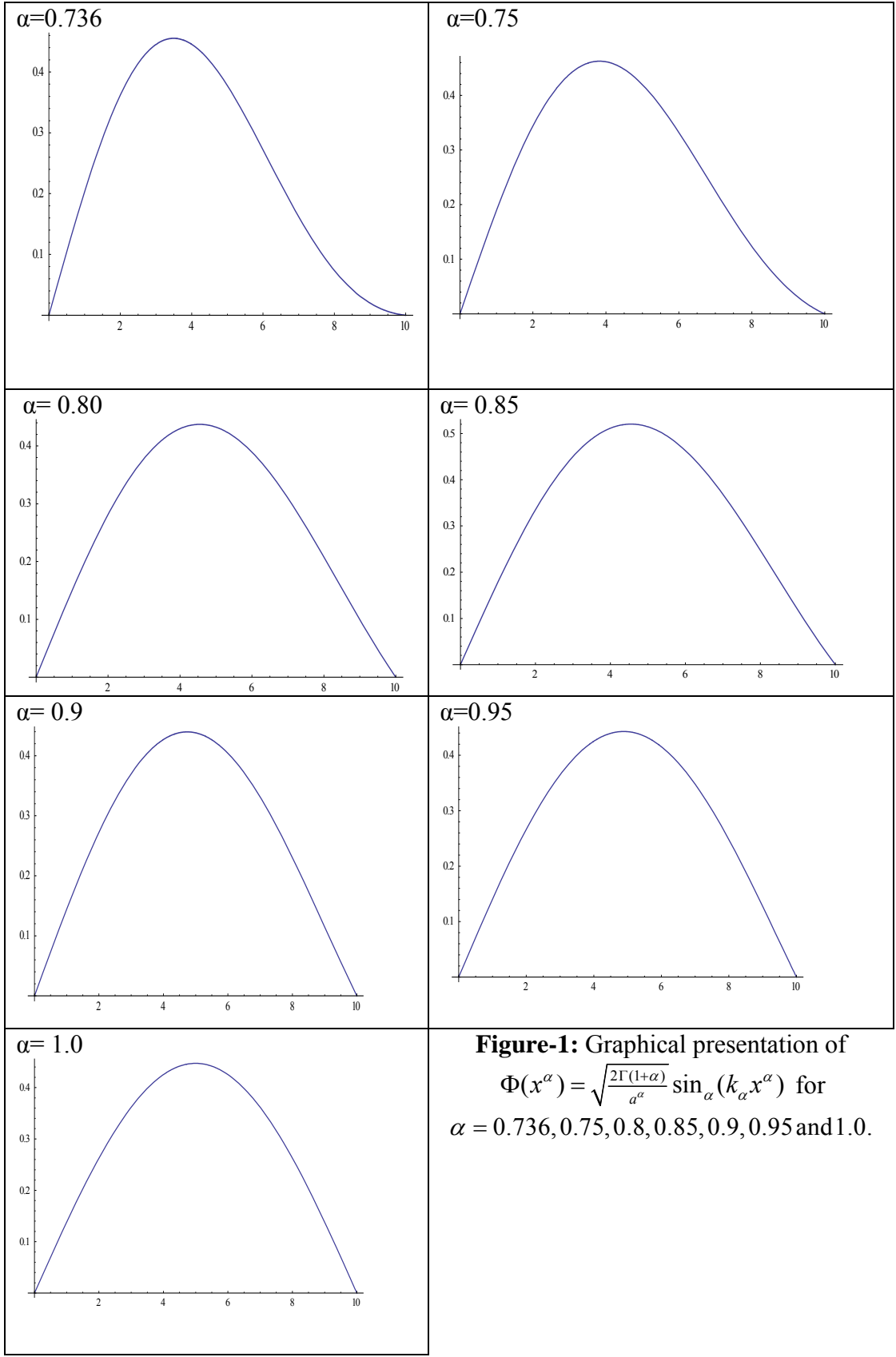

**Figure-1:** Graphical presentation of $\Phi(x^\alpha) = \sqrt{\frac{2\Gamma(1+\alpha)}{a^\alpha}} \sin_\alpha(k_\alpha x^\alpha)$ for $\alpha = 0.736, 0.75, 0.8, 0.85, 0.9, 0.95 \text{ and } 1.0$.

The plot is drawn $\Phi(x^\alpha) = \sqrt{\frac{2\Gamma(1+\alpha)}{a^\alpha}} \sin_\alpha(k_\alpha x^\alpha)$ against $x$. Here the box width is taken as 10 unit that is $a=10$. From numerical analysis we found that the quantum boundary conditions are satisfied up to $\alpha \approx 0.736$ from $\alpha = 1$. The plot suggests that the maxima of the wave function shifts to the right with the increase of $\alpha$ value. More than that the nature of wave function changes with $\alpha$ value. But at $\alpha = 1$ the plot is same as suggested by one dimensional Schrödinger potential box problem. At $\alpha \approx 0.736$, the curve is not symmetrical and more area covers in the left side than other. Less the $\alpha$ value means more asymmetrical is the plot. These plots have one zero crossing [16]. This means the quantum number of the system is 1. The system is in ground state. To compare the wave functions for various $\alpha$ values we have another plot which is given below-(Figure-2)

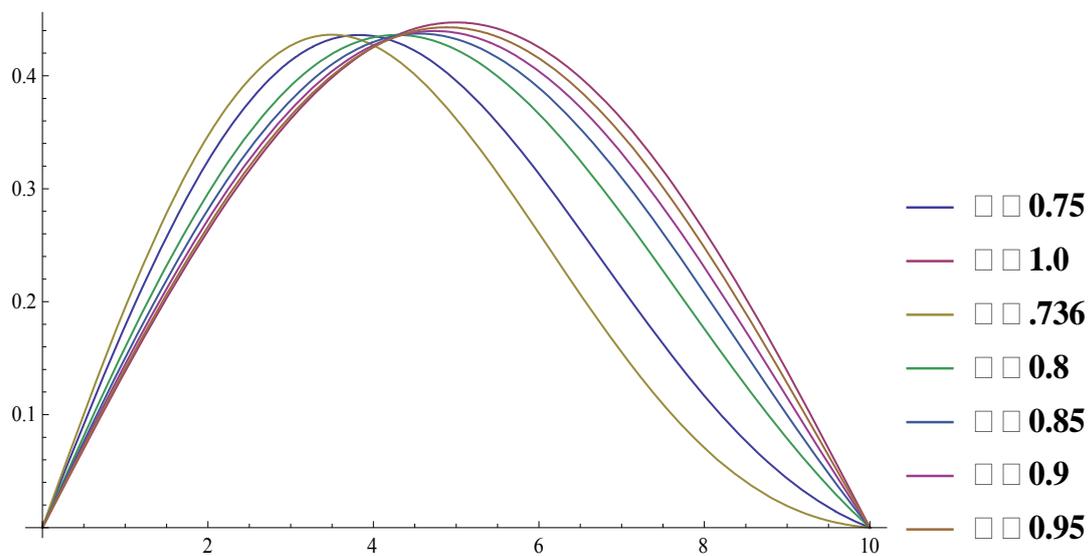

**Figure-2:** Graphical presentation of $\Phi(x^\alpha)$ for different values of $\alpha$ for a=10.

If we choose the box length as 5.7 units, the plot will be as below (Figure-3)

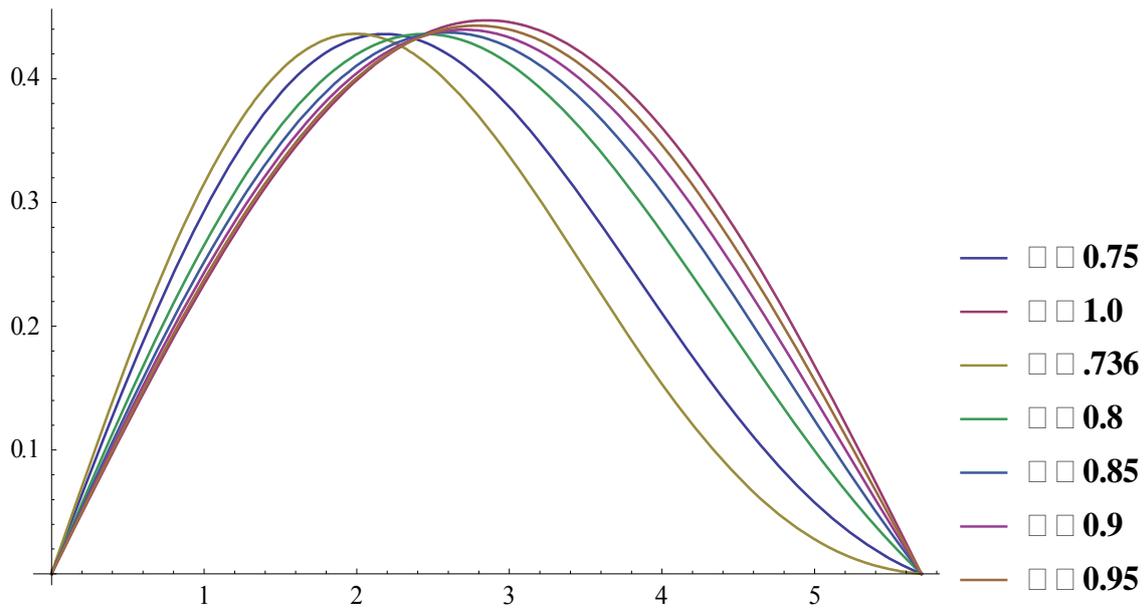

**Figure-3:** Graphical presentation of $\Phi(x^\alpha)$ for different values of $\alpha$ for a=5.7.

## 7.3 Probability density

As we got wave functions for various values of $\alpha$, we can also get probability density $\rho_\alpha = \psi_\alpha^* \psi_\alpha$. Probability density plot for various $\alpha$ is given below

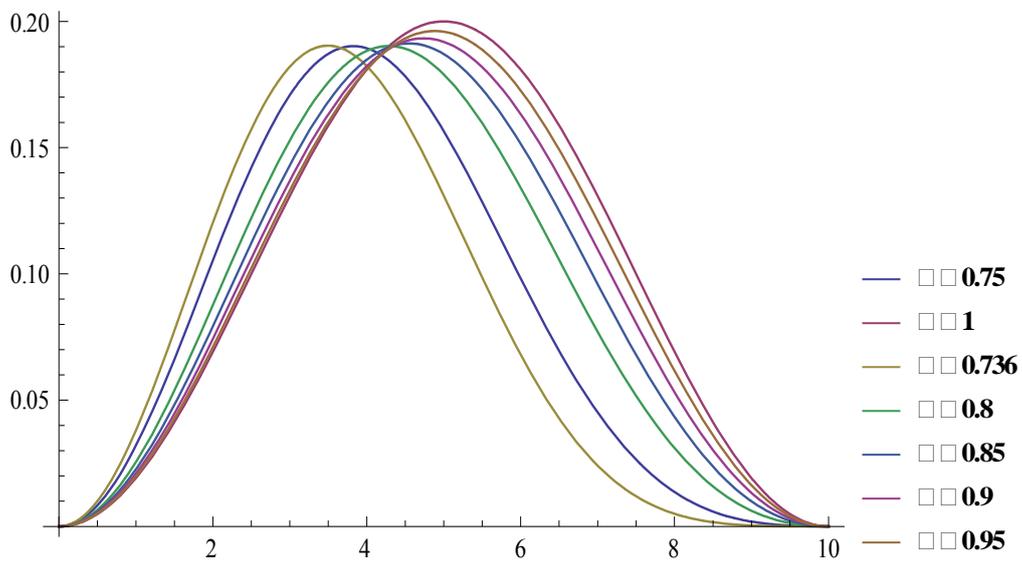

**Figure-4:** Graphical presentation of $\psi_\alpha^* \psi_\alpha$ for different values of $\alpha$ for a=10.

## 7.4 Energy calculation

Now we can calculate energy of the particle. Using equations (31a) and (34c) that is $(2^\alpha m_\alpha \varepsilon_\alpha)/\hbar_\alpha^2 = (M_\alpha/a)^{2\alpha}$ implying $\varepsilon_\alpha = (\hbar_\alpha^2/2^\alpha m_\alpha)(M_\alpha/a)^{2\alpha}$. For $\alpha = 1$ the energy is $\varepsilon_1 = (\hbar^2/2m)(\pi/a)^2$. This is the energy for first quantum state as described in quantum mechanics.

## 8.0 Conclusions

Using fractional derivative of Jumarie type we found that quantum mechanics in fractional region i.e. $0.736 < \alpha \leq 1$ region quantum behaviour of the particle changes dramatically. In this region equation of continuity is successfully maintained and the stationary condition also holds. For $\alpha = 1$ all the equations are normal classical the Schrödinger equation in normal space. We studied particle in a box problem and found that the wave equation in fractional sense also meets the condition. Further the wave function is not symmetric till $\alpha = 1$. For $0.736 < \alpha < 1$ the peak of wave function is left sided i.e. the peak is on the left side of the middle point of the box. The existence amplitude i.e. wave function has higher amplitude for higher $\alpha$ and it is maximum when $\alpha = 1$. Thus the wave function or existence amplitude is $\alpha$ dependent. We need further study on the fractional Quantum mechanics of $\alpha < 0.736$ to understand the internal behaviour of the quantum states.

# Appendix

## 1. Fractional mass

Fractional mass $m_\alpha$ may be defined as $m_\alpha = \rho \int dx^\alpha$, where $\rho$ is fractional linear mass density in one dimension. We have considered that the density of mass is same as it is in the case $\alpha = 1$.

## 2. Fractional velocity

The change of fractional displacement $\vec{dx^\alpha}$ per unit change in fractional time $dt^\alpha$ is the fractional velocity i.e. $\vec{v}_\alpha = \dfrac{\vec{d^\alpha x}}{dt^\alpha}$

## 3. Fractional wave length

Fractional wave length can be demonstrated by a plot (Figure: A-1) of a fractional wave of the order $\alpha = 0.8$

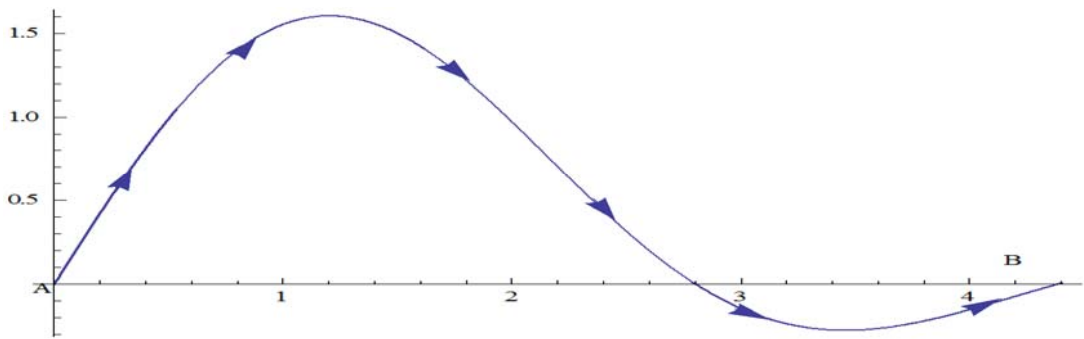

**Figure A1:** Showing fractional wave-length

The wave length is the distance $AB$. That is the distance covered by a fractional wave in a full fractional cycle. Fractional wave length is not a fixed quantity. It changes with the evolution of fractional time.

## 4. Fractional Time period

The time taken $N_\alpha$ to a wave to cover the distance *AB* is the fractional time period. We should take care that this is first order time period. As wavelength changes, the time period also changes with the wave propagation. But we assume that

$$\lambda_\alpha = v_\alpha N_\alpha$$

## 5. Fractional angular frequency

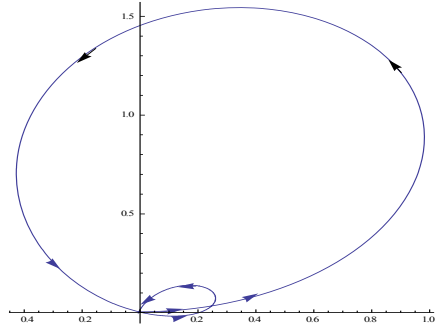

**Figure A2:** Showing concept of fractional angular frequency

The above plot is the Polar plot of the fractional wave of the order $\alpha = 0.8$. In this polar plot we can easily see that the wave is returned to the same point after completing a fractional cycle i.e. in its origin. By polar plot we can say that the angle traversed in a full fractional cycle is $2\pi$. Thus fractional angular frequency can be assigned as

$$\omega_\alpha = \frac{2\pi}{N_\alpha}$$

From this we can see that the product of fractional angular momentum and fractional time period $N_\alpha$ is always $2\pi$ though both are varying.

## 6. Fractional wave constant (or vector in 3Dimension)

From the analysis of fractional wave of equation (8) if $k_\alpha x^\alpha - \omega_\alpha t^\alpha = 0$ that is phase part of the wave is zero, we can get $k_\alpha = \omega_\alpha \left(t^\alpha / x^\alpha\right) = \omega_\alpha / v_\alpha$ as $x^\alpha / t^\alpha = v_\alpha$ is fractional velocity. Now $k_\alpha = \left(\omega_\alpha / v_\alpha\right) = 2\pi / \left(v_\alpha N_\alpha\right)$, using appendix (5). Now using appendix (4) we have

$$k_\alpha = 2\pi / \lambda_\alpha$$

## 7. Fractional reduced Plank constant

In this paper we have introduced fractional Plank constant $\hbar_\alpha$ as a basic constant for $\alpha$ ordered fractional system. For the limiting condition of $\alpha$ this constant is of the form of reduced Plank constant $\hbar$.

## 8. Theorem

If $f(x,y)$ be a function which is fractionally $\alpha$-th order differentiable with respect to both the variable $x$ and $t$ then $D_y^\alpha D_x^\alpha f(x,y) = D_x^\alpha D_y^\alpha f(x,y)$ or $f_{xy}^{2\alpha}(x,y) = f_{yx}^{2\alpha}(x,y)$ equivalently where $0 \leq \alpha \leq 1$.

**Proof:**

Consider a function $\phi(x) = f(x, y+k) - f(x,y)$, $k > 0$. Now fractional mean value theorem states that [25]

$$\phi(x+h) - \phi(x) = \frac{h^\alpha}{\Gamma(1+\alpha)} \phi_x^\alpha(x+\theta h), \text{ where } 0 < \theta < 1, 0 \leq \alpha \leq 1$$

$$\phi(x+h) - \phi(x) = \frac{h^\alpha}{\Gamma(1+\alpha)} [f_x^\alpha(x+\theta h, y+k) - f_x^\alpha(x+\theta h, y)]$$

Let $F(y) = f_x^\alpha(x+\theta h, y)$. Then using fractional Mean value theorem we have

$$\phi(x+h) - \phi(x) = \frac{h^\alpha}{\Gamma(1+\alpha)} [F(y+k) - F(y)] = \frac{h^\alpha}{\Gamma(1+\alpha)} \frac{k^\alpha}{\Gamma(1+\alpha)} [F_y^\alpha(y+\theta_1 k) - F_y^\alpha(y)]$$

$$= \frac{h^\alpha}{\Gamma(1+\alpha)} \frac{k^\alpha}{\Gamma(1+\alpha)} [f_{yx}^\alpha(x+\theta h, y+\theta_1 k)] \text{ where } 0 < \theta, \theta_1 < 1, 0 \leq \alpha \leq 1$$

On the other hand $\phi(x+h) = f(x+h, y+k) - f(x+h, y)$. Therefore

$$\phi(x+h) - \phi(x) = f(x+h, y+k) - f(x+h, y) - f(x, y+k) + f(x,y)$$

$$f_y^\alpha(x,y) = \Gamma(1+\alpha) \lim_{k \to 0} \frac{f(x, y+k) - f(x,y)}{k^\alpha}$$

or

$$f_{xy}^{2\alpha}(x,y) = \Gamma(1+\alpha) \lim_{h \to 0} \frac{f_y^\alpha(x, y+h) - f_y^\alpha(x,y)}{h^\alpha}$$

$$= (\Gamma(1+\alpha))^2 \lim_{h \to 0} \lim_{k \to 0} \frac{f(x+h, y+k) - f(x+h, y) - f(x, y+k) + f(x,y)}{k^\alpha h^\alpha}$$

$$= (\Gamma(1+\alpha))^2 \lim_{h \to 0} \lim_{k \to 0} \frac{\phi(x+h) - \phi(x)}{k^\alpha h^\alpha}$$

$$= (\Gamma(1+\alpha))^2 \lim_{h \to 0} \lim_{k \to 0} f_{yx}^{2\alpha}(x+\theta h, y+\theta_1 k)$$

or

$$f_{xy}^{2\alpha}(x,y) = f_{yx}^{2\alpha}(x,y)$$

Hence the theorem is proved. Which we have used in our detailed derivation.